# Adsorption properties of dopamine derivatives using carbon nanotubes: a first-principles study

Heeju Kim[1] and Gunn Kim[2] *

[1]Department of Physics, Sejong University, Seoul 05006, Republic of Korea

[2]Department of Physics & Astronomy and Institute of Fundamental Physics, Sejong University, Seoul 05006, Republic of Korea

*Corresponding author: Gunn Kim (gunnkim@sejong.ac.kr)

## Abstract

Detecting dopamine is of great biological importance because the molecule plays many roles in the human body. For instance, lack of dopamine release is the cause of Parkinson's disease. Although many researchers have carried out experiments on dopamine detection using carbon nanotubes (CNTs), there are only a few theoretical studies on this topic. We study the adsorption properties of dopamine and its derivatives, L-DOPA and dopamine o-quinone, adsorbed on a semiconducting (10, 0) CNT, using density functional theory calculations. Our computational simulations reveal that localized states originating from dopamine o-quinone appear in the bandgap of the (10, 0) CNT, but those originating from dopamine and L-DOPA do not appear in the gap. Therefore, dopamine o-quinone is expected to be detectable using an external electric field but dopamine and L-DOPA should be difficult to detect.

# 1. Introduction

Dopamine is one of the most fundamental neurotransmitters that act as messengers for synaptic transmission in a brain. It plays roles in motor control, attention, motivation, learning, reward, etc[1-4]. Because of its various roles in the human body, dopamine may cause serious mental illnesses. In particular, the addictions such as drug, gambling, and computer game addictions have relevance to the excessive release of dopamine and addictions arise from the repetition of behaviors that release dopamine[5-7]. Recently, some researchers have reported that dopamine is related to attention deficit/hyperactivity disorder[8-9]. Moreover, the lack of dopamine release is the cause of Parkinson's disease. Unfortunately, dopamine itself cannot pass through the blood-brain barrier which means that it cannot be administered therapeutically, L-DOPA, a precursor of dopamine, is used to increase the concentration of dopamine in the treatment of Parkinson's disease because it is able to pass through the barrier[10]. In this regard, monitoring the concentrations of dopamine and L-DOPA are of great biological significance. However, despite its importance in the brain, it is hard to detect the existence of dopamine. An efficient way to determine whether dopamine exists is by cyclic voltammetry (CV)[11-12]. In an experiment using a CV, dopamine is oxidized to dopamine o-quinone, and this, in turn, is reduced to dopamine again at the electrodes, by controlling the direction of the current[13]. When dopamine is oxidized to dopamine o-quinone, two hydrogen atoms in the hydroxyl groups are removed. The organic compound becomes highly reactive and may be toxic[14-17]. Therefore, we also consider dopamine o-quinone for this computational study.

Carbon nanotubes (CNTs) are attractive macromolecules which are applicable to the sensing of biomolecules because of their unique electrical and chemical properties[18-23]. CNTs can improve the electrochemical reactivity of biomolecules and alleviate surface fouling effects[24-25]. In this respect, CNT-based electrodes are commonly used in biological and CV experiments[26-27]. Many researchers have already done various experimental investigations into the detection of dopamine, L-DOPA, and dopamine o-quinone[28-30]. There are a few papers on the dopamine adsorption on the CNT[31-32]. However, the literature does not compare dopamine with L-DOPA and dopamine o-quinone for the adsorption on the CNT. As mentioned above, there are strong motivations to examine the properties of L-DOPA and dopamine o-quinone and gain theoretical insights into dopamine biochemistry.

In this paper, we report a first-principles investigation of the adsorption properties of dopamine, L-DOPA, and dopamine o-quinone molecules adsorbed on the semiconducting (10, 0) CNT. The binding energies of dopamine, L-DOPA, and dopamine o-quinone were 0.69, 0.69, and 0.75 eV, respectively. Our results show that the localized states originating from dopamine and L-DOPA do not exist in the bandgap of the tube, but that those from dopamine o-quinone appear in the bandgap.

## 2. Computational Details

Based on density functional theory (DFT)[33], we have investigated CNTs with respect to the adsorption of dopamine, L-DOPA, and dopamine o-quinone. Using the Quantum ESPRESSO package[34], wave eigenfunctions and eigenenergies were calculated, using a plane-wave basis set to kinetic cutoff energy of 40 Ry (~540 eV). Ion–electron interactions were described with projector augmented wave (PAW) pseudopotentials[35]. For the exchange-correlation part of the Hamiltonian, we used Perdew–Burke–Ernzerhof parameterization of the generalized gradient approximation method[36]. To get more accurate results, we also considered van Der Waals correction with Grimme's D2 method (DFT-D2)[37]. The estimated energy error was less than $10^{-7}$ and the convergence threshold on forces for ionic minimization was 0.03 eV/Å. Three molecules — dopamine, L-DOPA, and dopamine o-quinones — were calculated, in a simple cubic lattice with a lattice constant of 20 Å. The $\Gamma$ point was used in the calculations of the molecules. For the host material, we selected the (10, 0) CNT. The unit cell size was 25.0 × 25 × 4.233 Å and the axis of the tube was in the $z$-direction. A $k$-point grid for computation of the CNT was $1 \times 1 \times 10$ as in a Monkhorst–Pack grid[38]. With respect to adsorption of molecules on the CNT, the initial locations of the molecules were 3.3 Å away from the tube wall. For each molecular adsorption, the binding energy was defined as $E_{binding} = E_{CNT+molecule} - (E_{CNT} + E_{molecule})$, where $E_{CNT+molecule}$ is the total energy of a molecule-adsorbed CNT, $E_{CNT}$ is the total energy of the bare CNT, and $E_{molecule}$ is the energy of the isolated molecule in a vacuum. For the stacking configuration, AB stacking between the catechol ring of each molecule and a CNT hexagon was considered.

## 3. Results and Discussion

Figure 1 shows model structures of (a) dopamine, (b) L-DOPA, and (c) dopamine o-quinone. Dopamine has a catechol ring and an amine group at the end of a linear hydrocarbon. Since dopamine has several isomers considering the orientation of two hydroxyl groups, we calculated a few different isomers.

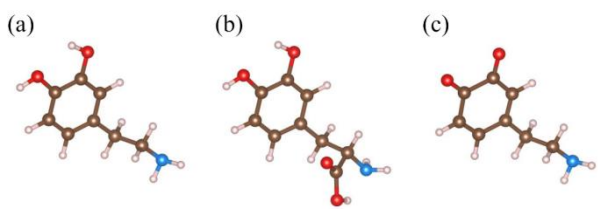

Figure 1. Optimized structures of (a) dopamine, (b) L-DOPA, and (c) dopamine o-quinone. Brown,

red, blue, and white spheres represent carbon, oxygen, nitrogen, and hydrogen atoms, respectively.

The three types of molecules have structural similarities, but there are notable differences in their electrical properties due to their different chemical groups. Compared with dopamine, L-DOPA contains a carboxyl group instead of a hydrogen atom on the linear carbon chain. For dopamine o-quinone, the hydrogen atoms are separated from the hydroxyl groups of dopamine, and the organic compound is in an unstable state. The differences among the electric dipole moments of the molecules are remarkable (Table 1); the dipole moment of dopamine o-quinone is approximately twice as much as those of the others. Furthermore, the energy gap between the highest occupied molecular orbital (HOMO) and the lowest unoccupied molecular orbital (LUMO) of dopamine o-quinone is the smallest among the three molecules. For dopamine and L-DOPA, the HOMO-LUMO gap is ~ 4 eV. Considering the fact that a molecule with a larger HOMO–LUMO gap is more stable, dopamine and L-DOPA molecules should be stable and dopamine o-quinone relatively unstable. Therefore, we can conclude that dopamine is similar to L-DOPA with respect to energy spacing, but it is very different from dopamine o-quinone.

Table 1. Electric dipole moments and HOMO-LUMO gaps of dopamine, L-DOPA, and dopamine o-quinone.

|  | Dopamine | L-DOPA | Dopamine o-quinone |
| --- | --- | --- | --- |
| Electric dipole moment (debye) | 2.98 | 3.35 | 6.75 |
| HOMO-LUMO gap (eV) | 4.01 | 3.94 | 1.11 |

The wave functions of the three types of molecules also show interesting features, as displayed in Figure 2. The wave functions of the HOMO and LUMO of dopamine and L-DOPA have similar shapes. In contrast, dopamine o-quinone has a different HOMO shape compared to those of dopamine and L-DOPA, while its LUMO is quite similar to the HOMO states of dopamine and L-DOPA. In addition, the HOMO-1 and LUMO of dopamine look similar to the HOMO-1 and LUMO+1 of dopamine o-quinone, respectively. This indicates that the HOMO of dopamine o-quinone forms as a new state when the two H atoms are removed from the two hydroxyl side groups of the catechol ring, leaving very unstable oxygen atoms. Thus the dangling bonds at the O atoms are associated with the HOMO and the LUMO of the dopamine molecule. When dopamine o-quinone is adsorbed on the CNT, we will pay attention to the HOMO state.

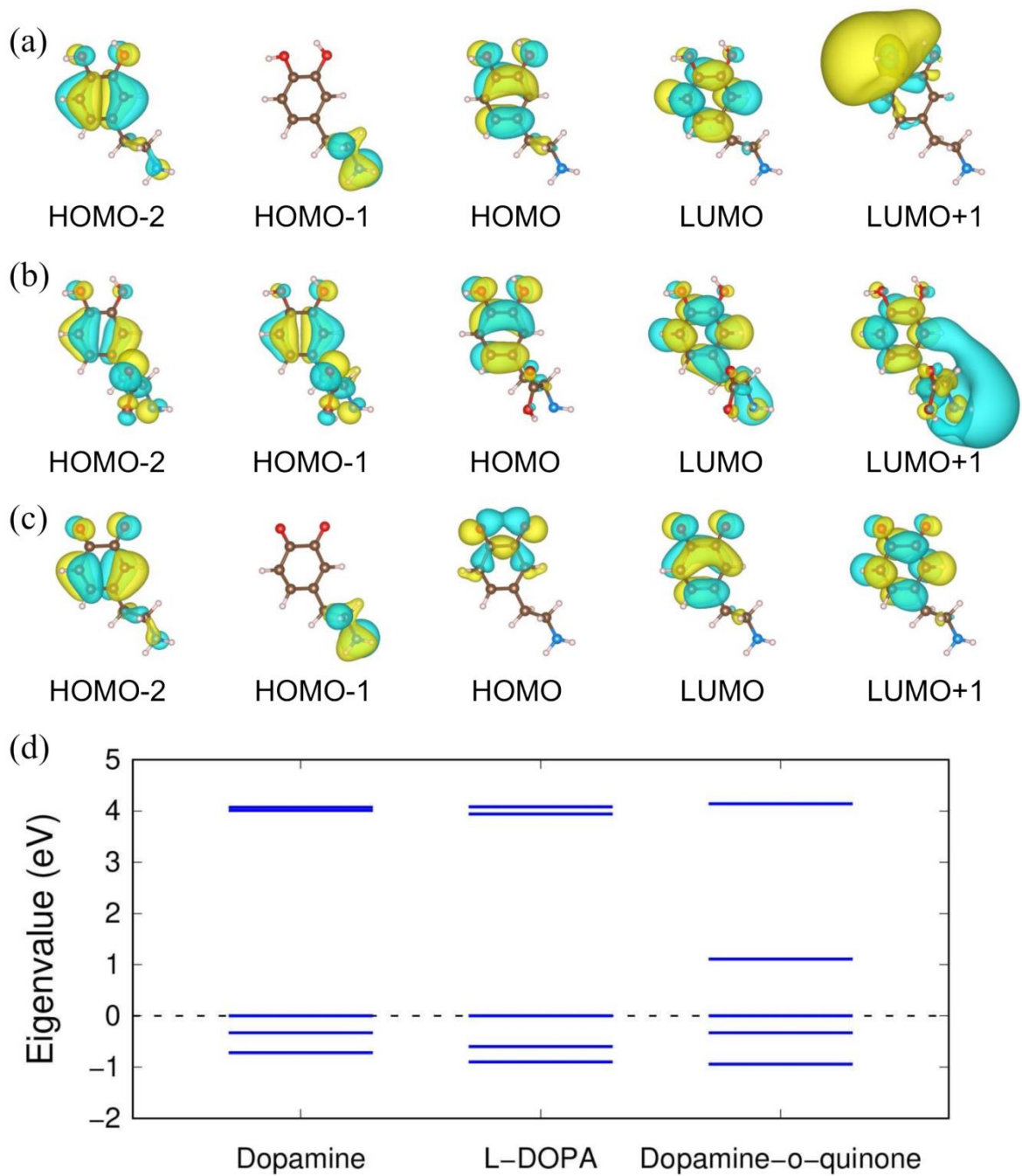

Figure 2. Energy eigenfunctions of (a) dopamine, (b) L-DOPA, (c) dopamine o-quinone molecule; the wave functions represent from HOMO-2 to LUMO+1 in order. The HOMO level of each molecule is set to zero in (d).

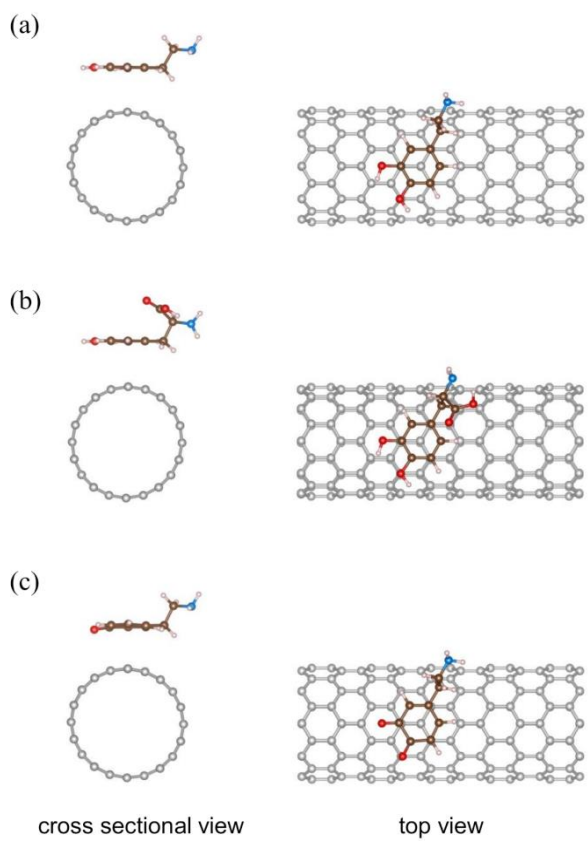

cross sectional view    top view

Figure 3. Model Structures of (a) dopamine, (b) L-DOPA, and (c) dopamine o-quinone adsorption. Brown spheres represent carbon atoms and red, blue, and white spheres indicate oxygen, nitrogen, and hydrogen atoms, respectively. In each case, the molecule is in AB stacking with respect to the CNT hexagon.

Next, we turn to the (10, 0) CNT, which has a bandgap of 0.75 eV in its pristine state[39] with an adsorbed molecule (dopamine, L-DOPA or dopamine o-quinone). Among several binding configurations, we determined the energetically most favorable configurations using first-principles calculations. For the molecules placed on graphene, the common trend is that AB stacking is a few meV more stable than AA stacking[40-41]. Thus, we present the adsorption configurations with AB stacking between the phenyl ring and the CNT in Figure 3. The catechol ring is in an AB stacking configuration with a hexagon of the nanotube. For the adsorption of dopamine, L-DOPA, and dopamine o-quinone on the CNT, the binding energies are summarized in Table 2. The adsorption energies of dopamine and L-DOPA on the (10, 0) CNT are the same (0.69 eV)[42]. For dopamine o-quinone, the binding energy is somewhat higher by 0.06 eV than dopamine and L-DOPA. As mentioned above, since the oxygen atoms on the catechol ring have dangling bonds, slightly stronger binding occurs when dopamine o-quinone is adsorbed on a CNT.

Table 2. Binding Energies of dopamine, L-DOPA, and dopamine o-quinone onto the (10, 0) CNT.

|  | Dopamine | L-DOPA | Dopamine o-quinone |
| --- | --- | --- | --- |
| Binding energy (eV) | 0.69 | 0.69 | 0.75 |

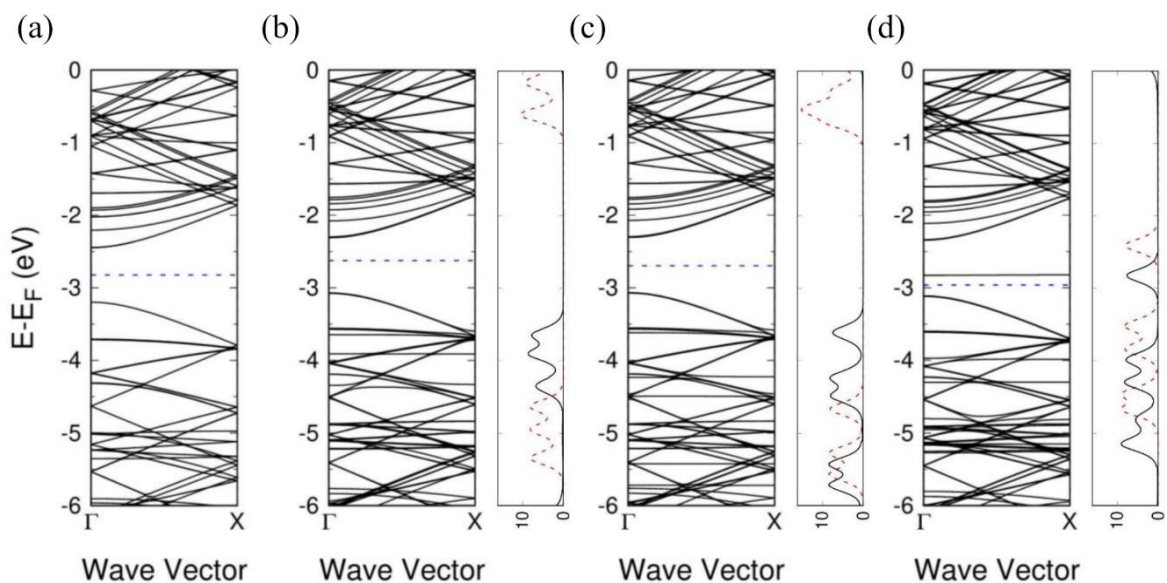

Figure 4. Band structures of (a) pristine (10, 0) CNT and (b) dopamine-, (c) L-DOPA- (d) dopamine o-quinone-adsorbed CNT. For the molecule-adsorbed CNTs, the PDOS of each molecule is also shown on the right side of each panel. In the PDOS, the red dotted lines indicate data for the molecules in a vacuum, and black solid ones represent for the PDOS after adsorption. The bandgap size is 0.75 eV for pristine CNT, and 0.76, 0.76 and 0.77 eV for dopamine-, L-DOPA-, dopamine o-quinone-adsorbed CNT, respectively. The Fermi levels are –2.82, –2.62, –2.70, and –2.96 eV for (a), (b), (c), and (d), respectively, which are marked by dotted lines.

In all the model systems we consider here, the bond lengths or angles of the molecules are almost the same as those in a vacuum. When dopamine or L-DOPA is adsorbed on the tube, there is no significant change in the bandgap of the tube. On the other hand, the flat bands are around –4 eV, as seen in Figure 4, which correspond to the HOMO of the adsorbed molecule. In the projected density of states (PDOS), the PDOS of dopamine after adsorption (the black line) is ~1 eV higher than that of dopamine in a vacuum (the red line). The upshift of the molecular energy level after adsorption means that electron donation occurs from the adsorbate to the CNT. In addition, this trend can be supported by the Fermi level changes. After dopamine molecule is adsorbed, the Fermi level is upshifted to –2.62 eV, comparing the pure CNT Fermi level of –2.82 eV, which evidently shows that the molecule is likely to donate electron to the CNT. Because L-DOPA has very similar electronic properties to those of dopamine around the Fermi level, the two band structures closely resemble each other. Interestingly, when dopamine o-quinone is adsorbed on the CNT, a remarkable feature is revealed. First, even though the bandgap of the CNT does not show a significant change, a flat band (an in-gap state), originating from the LUMO state of dopamine o-quinone, appears 0.13 eV above the Fermi level, as shown in Figure 4(d). Compared with the PDOS of dopamine o-quinone in a vacuum, the PDOS of the molecule on the CNT is ~0.5 eV lower, which indicates that the dopamine o-quinone states are downshifted by adsorption. Unlike the two molecules, when dopamine o-quinone molecule is adsorbed on CNT, the Fermi level is downshifted to –2.96 eV, confirming the molecule tends to accept electron from the CNT. Thus, it is expected that dopamine o-quinone acts as an acceptor and the tube becomes a weak p-type semiconductor. If we use a CNT-based transistor for the molecule detection, the current-voltage (*I-V*) characteristics will vary with doping and thus the CNT could act as a sensor for dopamine o-quinone. When molecules are adsorbed on the transistor, the localized states of the adsorbates act as scatterers near the Fermi level and result in the changes in the *I-V* curve of the transistor. We also tried vertical adsorption of dopamine o-quinone on CNT to see what happens if the LUMO of the molecule approach to the CNT. This system is unstable and the energy of 1.05 eV is required to make this endothermic reaction occur.

For the case of the adsorption of dopamine o-quinone on the CNT, further analysis of the localized state in the bandgap should prove meaningful to understand. We applied an external electric field with the magnitude of 0.1V/Å perpendicular to the tube axis. In the computation, the tube axis is in the *z* direction and the field is in ±*y* directions. For an electric field in the +*y* direction, the bandgap of the CNT was the same as the one before adsorption which is calculated to be 0.76 eV. However, the binding energy was increased by 0.13 eV owing to the electrostatic attraction. In addition, the localized states of dopamine o-quinone are upshifted by 0.27 eV, compared to the case in the absence of the electric field. It means that dopamine o-quinone is likely to donate electron to the CNT. Figure 5(a) shows the electronic structure of the CNT with dopamine o-quinone. Compared with Fig. 4(d), the flat band derived from dopamine o-quinone is relatively far from the valence band maximum. The dotted and solid lines in the PDOS indicate the LUMO states of dopamine o-quinone in the presence and absence of the applied electric field, respectively. It is obvious that the LUMO of dopamine o-quinone moves away from the Fermi level and goes up when the electric field is applied in the +*y* direction. In Fig. 5(b), yellow (cyan) represents electron accumulation (depletion) formed by the charge density rearrangement upon the application of the external electric field. When the electric field is applied in the –y direction, on the other hand, the system is unstable and the potential is not converged easily. This is because the electric field makes the LUMO of dopamine o-quinone partially filled as the state in the bandgap approaches the Fermi level. In this case, the LUMO state is downshifted by a few meV. The binding energy of the adsorbate is almost the same as the case without an electric field.

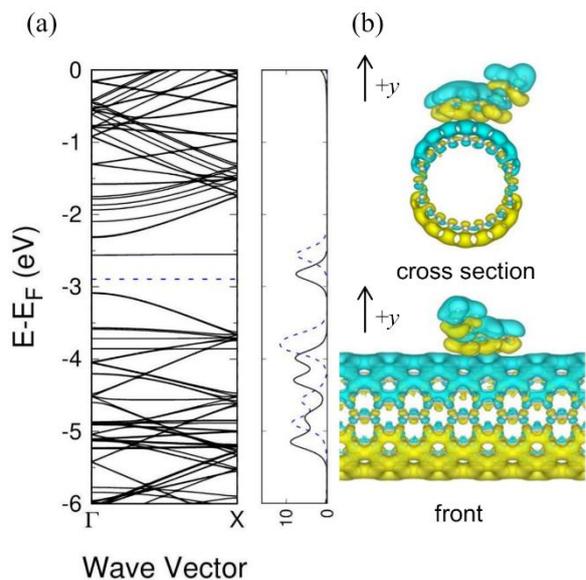

Figure 5. (a) Band structure of the CNT when dopamine o-quinone molecule is adsorbed in the presence of an external electric field. The Fermi level is −2.87 eV for this system. On the right of the panel, the dotted and solid lines indicate the localized states of dopamine o-quinone in the presence and absence of the applied electric field, respectively. (b) Charge density difference plots for the dopamine o-quinone-adsorbed (10, 0) CNT under the electric field in the +y direction; yellow and cyan present electron accumulation and depletion, respectively.

## 4. Conclusion

We performed first-principles calculations to investigate the adsorption properties of dopamine, L-DOPA, and dopamine o-quinone molecules on CNT. Considering various adsorption configuration, we found that AB stacking between the catechol rings and the CNTs' hexagon was the most stable. The binding energies we obtained were 0.69, 0.69, and 0.75 eV for dopamine, L-DOPA, and dopamine o-quinone but the bandgap of the CNT and the geometries of the molecules did not significantly change upon adsorption. Dopamine and L-DOPA tend to donate electrons to the CNT as localized states in the band structures showing upshifting tendencies. Unlike dopamine and L-DOPA, dopamine o-quinone accepts electrons from the CNT and a flat band occurs within the bandgap. When a uniform electric field is applied to this system in the +$y$ direction, dopamine o-quinone donates electrons which it accepted from the CNT under no external electric field. This property is evidence that dopamine o-quinone could be detected by CNTs in experiments whereas dopamine and L-DOPA will be hard to detect.

## Acknowledgments

This work was supported by the Mid-career Researcher Program [Grant No. 2016R1A2B2016120] through the National Research Foundation funded by the Korea government.

## Reference

[1] R. A.Wise, Dopamine, learning and motivation, Nat. Rev. Neurosci. 5 (2004) 483–494. https://doi.org/10.1038/nrn1406

[2] K. C. Berridge, T. E. Robinson, J. W. Aldridge, Dissecting components of reward: 'liking', 'wanting', and learning. Curr. Opin. Pharmacol. 9 (2009) 65–73. https://doi.org/10.1016/j.coph.2008.12.014

[3] T. T-J. Chong, Updating the role of dopamine in human motivation and apathy. Curr. Opin. Behav. Sci. 2 (2018), 35–41. https://doi.org/10.1016/j.cobeha.2017.12.010

[4] D. Meder, D. M. Herz, J. B. Rowe, S. Lehéricy, H. R. Siebner, The role of dopamine in the brain - lessons learned from Parkinson's disease. NeuroImage. https://doi.org/10.1016/j.neuroimage.2018.11.021 (accessed Nov 20, 2018)

[5] ND Volkow, JS Fowler, G-J Wang, JM Swanson, Dopamine in drug abuse and addiction: results from imaging studies and treatment implications. Mol. Psychiatry 9 (2004) 557–569. https://doi.org/10.1038/sj.mp.4001507

[6] I. H.A. Franken, J. Booij, W. van der Brink, The role of dopamine in human addiction: From reward to motivated attention. Eur. J. Pharmacol. 526 (2005) 199-206. https://doi.org/10.1016/j.ejphar.2005.09.025

[7] R. J. van Holst, G. Sescousse, L. K. Janssen, M. Janssen, A. S. Berry, W. J. Jagust, R. Cools, Increased Striatal Dopamine Synthesis Capacity in Gambling Addiction. Biol. Psychiatry 83 (2018) 1036-1043. https://doi.org/10.1016/j.biopsych.2017.06.010

[8] J. Biederman, Attention-Deficit/Hyperactivity Disorder: A Selective Overview. Biol. Psychiatry 57 (2005) 1215-1220. https://doi.org/10.1016/j.biopsych.2004.10.020

[9] N. del Campo, S. R. Chamberlain, B. J. Sahakian, T. W. Robbins, The Roles of Dopamine and Noradrenaline in the Pathophysiology and Treatment of Attention-Deficit/Hyperactivity Disorder. Biol. Psychiatry 69 (2011) e145-e157. https://doi.org/10.1016/j.biopsych.2011.02.036

[10] R. Cabezas, M. Ávila, J. Gonzalez, R. S. El-Bachá, E. Báez, L. M. García-Segura, J. C. J. Coronel, F. Capani, G. P. Cardona-Gomez, G. E. Barreto, Astrocytic modulation of blood brain barrier: perspectives on Parkinson's disease. Front. Cell. Neurosci. 8 (2014), Article 211. https://doi.org/10.3389/fncel.2014.00211

[11] P. T. Kissinger, W. R. Heineman, Cyclic voltammetry. J. Chem. Educ. 60 (1983) 702-706. https://doi.org/10.1021/ed060p702

[12] D. L. Robinson, B. J. Venton, M. L.A.V. Heien, R. B. Wightman, Detecting Subsecond Dopamine Release with Fast-Scan Cyclic Voltammetry in Vivo. Clin. Chem. 49 (2003) 1763–1773. https://doi.org/10.1373/49.10.1763

[13] P. J. Britto, K. S. V. Santhanam, P. M. Ajayan, Carbon nanotube electrode for oxidation of dopamine. Bioelectrochem. Bioenerg. 41 (1996) 121-125. https://doi.org/10.1016/0302-4598(96)05078-7

[14] D. Ben-Shachar, R. Zuk, Y. Glinka, Dopamine Neurotoxicity: Inhibition of Mitochondrial Respiration, J. Neurochem. 64 (1995) 718-723. https://doi.org/10.1046/j.1471-4159.1995.64020718.x

[15] TG Hastings, DA Lewis, MJ Zigmond, Role of oxidation in the neurotoxic effects of intrastriatal dopamine injections, PNAS 93 (1996) 1956-1961. https://doi.org/10.1073/pnas.93.5.1956

[16] F. Solano, V. J. Hearing, J. C. García-Borrón, Neurotoxicity due to o-Quinones: Neuromelanin formation and possible mechanisms for o-Quinone detoxification, Neurotoxic. Res. 1 (1999) 153–169. https://doi.org/10.1007/BF03033287

[17] J. L. Bolton, M. A. Trush, T. M. Penning, G. Dryhurst, T. J. Monks, Role of Quinones in Toxicology, Chem. Res. Toxicol. 13 (2000) 135-160. https://pubs.acs.org/doi/abs/10.1021/tx9902082

[18] L. Chico, V. H. Crespi, L. X. Benedict, S. G. Louie, M. L. Cohen, Pure Carbon Nanoscale Devices: Nanotube Heterojunctions. Phys. Rev. Lett. 76 (1996) 971–974. https://doi.org/10.1103/PhysRevLett.76.971

[19] S. J. Tans, A. R. M. Verschueren, C. Dekker, Room-temperature transistor based on a single carbon nanotube. Nature 393 (1998) 49–52. https://doi.org/10.1038/29954

[20] C. Zhou, J. Kong, E. Yenilmez, H. Dai, Modulated Chemical Doping of Individual Carbon Nanotubes. Science 290 (2000) 1552–1555. https://doi.org/10.1126/science.290.5496.1552

[21] J.-C. Charlier, X. Blase, S. Roche, Electronic and transport properties of nanotubes. Rev. Mod. Phys. 79 (2007) 677–732. https://doi.org/10.1103/RevModPhys.79.677

[22]    W. I. Choi, J. Ihm, G. Kim, Modification of the electronic structure in a carbon nanotube with the charge dopant encapsulation. Appl. Phys. Lett. 92 (2008) 1–3. https://doi.org/10.1063/1.2929381

[23]    H. Wang, Y. Liu, S. Yao, G. Hu, Fabrication of super pure single−walled carbon nanotube electrochemical sensor and its application for picomole detection of olaquindox. Anal. Chim. Acta 1049 (2019) 82-90. https://doi.org/10.1016/j.aca.2018.10.024

[24]    J. Wang, Carbon-Nanotube Based Electrochemical Biosensors: A Review. Electroanalysis 17 (2005) 7-14. https://doi.org/10.1002/elan.200403113

[25]    W. Harreither, R. Trouillon, P. Poulin, W. Neri, A. G. Ewing, G. Safina, Carbon Nanotube Fiber Microelectrodes Show a Higher Resistance to Dopamine Fouling. Anal. Chem. 85 (2013) 7447-7453. https://doi.org/10.1021/ac401399s

[26]    B. L. Allen, P. D. Kichambare, A. Star, Carbon Nanotube Field-Effect-Transistor-Based Biosensors. Adv. Mater. 19 (2007) 1439–1451. https://doi.org/10.1002/adma.200602043

[27]    G. A. Rivas, M. D. Rubianes, M. C. Rodríguez, N. F. Ferreyra, G. L. Luque, M. L. Pedano, S. A. Miscoria, C. Parrado, Carbon nanotubes for electrochemical biosensing. Talanta 74 (2007) 291–307. https://doi.org/10.1016/j.talanta.2007.10.013

[28]    A. L. Sanati, F. Faridbod, M. R. Ganjali, Synergic effect of graphene quantum dots and room temperature ionic liquid for the fabrication of highly sensitive voltammetric sensor for levodopa determination in the presence of serotonin. J. Mol. Liq. 241 (2007) 316–320. https://doi.org/10.1016/j.molliq.2017.04.123

[29]    X. Ren, J. Ge, X. Meng, X. Qiu, J. Ren, F. Tang, Sensitive detection of dopamine and quinone drugs based on the quenching of the fluorescence of carbon dots. Sci. Bull. 61 (2016) 1615–1623. https://doi.org/10.1007/s11434-016-1172-1

[30]    J. Lin, B. Huang, Y. Dai, J. Wei, Y. Chen, Chiral ZnO nanoparticles for detection of dopamine. Mater. Sci. Eng., C 93 (2018) 739–745. https://doi.org/10.1016/j.msec.2018.08.036

[31]    M. Hesabi, The Interaction between Dopamine and Carbon Nanotube: A DFT and NBO Approach. J. Phys. Theor. Chem. IAU Iran. 8 (2012) 261-265.

[32]    M. JALILI, First Principal and QM/MM study of Dopamine Adsorption on Single Wall Carbon Nano Tubes and Single Wall Boroan Nitride Nano Tubes. Orient. J. Chem. 32 (2016) 1589-1600. http://dx.doi.org/10.13005/ojc/320335

[33]    W. Kohn, L. J. Sham, Self-Consistent Equations Including Exchange and Correlation Effects. Phys. Rev. 140 (1965) A1133−A1138. https://doi.org/10.1103/PhysRev.140.A1133

[34]    P. Giannozzi, S. Baroni, N. Bonini, M. Calandra, R. Car, C. Cavazzoni, D. Ceresoli, G. L. Chiarotti, M. Cococcioni, I. Dabo, A. D. Corso, S. Gironcoli, S. Fabris, G. Fratesi, R. Gebauer, U. Gerstmann, C. Gougoussis, A. Kokalj, M. Lazzeri, L. Martin-Samos, N. Marzari, F. Mauri, R. Mazzarello, S. Paolini, A. Pasquarello, L. Paulatto, C. Sbraccia, S. Scandolo, G. Sclauzero, A. P. Seitsonen, A. Smogunov, P. Umari, R. M. Wentzcovitch, QUANTUM ESPRESSO: a modular and open-source software project for quantum simulations of materials. J. Phys.: Condens. Matter. 21 (2009) 335502.

[35]     G. Kresse, D. Joubert, From ultrasoft pseudopotentials to the projector augmented-wave method. Phys. Rev. B: Condens. Matter Mater. Phys. 59 (1999) 1758−1775. https://doi.org/10.1103/PhysRevB.59.1758

[36]     J. P. Perdew, K. Burke, M. Ernzerhof, Generalized Gradient Approximation Made Simple. Phys. Rev. Lett. 77 (1996) 3865−3868. https://doi.org/10.1103/PhysRevLett.77.3865

[37]     S. Grimme, Semiempirical GGA-type density functional constructed with a long-range dispersion correction. J. Comput. Chem. 27 (2006) 1787–1799. . https://doi.org/10.1002/jcc.20495

[38]     H. J. Monkhorst, J. D. Pack, Special Points for Brillouin-Zone Integrations. Phys. Rev. B 13 (1976) 5188−5192. https://doi.org/10.1103/PhysRevB.13.5188

[39]     Y. Lee, D. Kwon, G. Kim, Y. Kwon, Ab initio study of aspirin adsorption on single-walled carbon and carbon nitride nanotubes. Phys. Chem. Chem. Phys. 19 (2017) 8076-8081. https://doi.org/10.1039/C6CP08122C

[40]     M. Birowska, K. Milowska and J. A. Majewski, Van Der Waals Density Functionals for Graphene Layers and Graphite. Acta Phys. Pol. A 120 (2011) 845-848. https://doi.org/10.12693/APhysPolA.120.845

[41]     J. Lee, K.-A. Min, S. Hong, G. Kim, Ab initio study of adsorption properties of hazardous organic molecules on graphene: Phenol, phenyl azide, and phenylnitrene. Chem. Phys. Lett. 618 (2015) 57-62. https://doi.org/10.1016/j.cplett.2014.10.064

[42]     C. Yeh, Y. Hsiao, J. Jiang, Dopamine sensing by boron and nitrogen co-doped single-walled carbon nanotubes: A first-principles study. Appl. Surf. Sci. 473 (2019) 59-64